\documentclass[aps,pra,preprint,superscriptaddress,showpacs]{revtex4}

\usepackage{amsmath,amssymb,amsfonts}
\usepackage{graphicx}
\usepackage{epsf}

\newcommand{\bra}[1]{\langle{#1}\vert}
\newcommand{\ket}[1]{\vert{#1}\rangle}
\newcommand{\fref}[1]{Fig.~\ref{#1}}

\begin{document}

\title{Phase properties of operator valued measures in phase space}

\author{T. Subeesh}
\affiliation{Dept. of Physics, Indian Institute of Technology Madras, Chennai 600036, India}
\affiliation{Dept. of Physics, Amrita School of Engineering, Amrita Vishwa Vidyapeetham, Ettimadai 641105, 
Coimbatore, India}
\email[]{t.subeesh@gmail.com}

\author{Vivishek Sudhir}

\affiliation{Dept. of Physics, Imperial College London, SW7 2AZ, UK}

\date{\today}

\begin{abstract}
The Wigner Phase Operator (WPO) is identified as an operator valued measure (OVM) and
its eigen states are obtained. An operator satisfying the canonical commutation relation
with the Wigner phase operator is also constructed and this establishes a Wigner 
distribution based operator formalism for the Wigner Phase Distribution. The operator satisfying 
the canonical commutation relation with the Wigner Phase Operator valued measure (WP-OVM)
is found to be not the usual number operator. We show a way to
overcome the non-positivity problem of the WP-OVM by defining a positive OVM by means
of a proper filter function, based on the view that phase measurements are 
coarse-grained in phase space, leading to the well known Q-distribution. The identification 
of Q phase operator as a POVM is in good agreement with the earlier observation regarding 
the relation between operational phase measurement schemes and the Q-distribution. The Q phase
POVM can be dilated in the sense of Gelfand-Naimark, to an operational setting of interference
at a beam-splitter with another coherent state - this results in a von Neumann projector with
well-defined phase.
\end{abstract}
\pacs{42.50.-p, 42.50.Dv, 03.67.Bg, 42.50.Ar}
\keywords{phase properties of quantized radiation field; wigner phase distribution; hermitian phase operator}

\maketitle

\section{Introduction}

The phase of a harmonic oscillator, and hence of the single-mode radiation field,
has proved to be a challenging question since the inception of quantum theory \cite{BarVac07, Nieto93}.

The earliest attempts at quantizing the phase, focused on ascribing a consistent meaning to the
polar decomposition of the field ladder operators \cite{Dirac27}. But such a phase operator suffers
from a pathological non-unitarity which can be traced back to the existence of a null-subspace, the one
spanned by the ground state, for the number operator \cite{Sten93}. Subsequent attempts relying essentially
on the polar decomposition overcomes this difficulty by introducing real trigonometric phase operator
\cite{Loui63,SusGlo64,CarNie65}, and phase difference operator \cite{CarNie68}.
Such a phase operator has been found to be useful in an operational
setting considered by Noh et al. \cite{NFM92}. The hermitian phase operator introduced by Pegg and Barnett
\cite{PegBar86, PegBar89}, relying on well-defined phase states in a finite dimensional Fock space, has
since proved to be definitive.

A complimentary, but intimately related \cite{BarnDalt93,SchBan92} approach to the quantum phase
is via the radial integral of the various quasi-probability distributions in phase
space \cite{GarKni92, TanGan93}. 
An operator intended to elucidate the phase structure inherent in the Wigner phase distribution
was constructed and studied in an earlier paper \cite{SubViv11}.

In \cite{SubViv11}, the operator corresponding to the radially integrated
Wigner distribution was shown to take the following diagonal form,
\begin{equation}\label{rhow_OVM}
    \hat{\rho}_\text{W}(\theta) = \frac{1}{\pi} \int_0^\infty du\; u e^{-u^2} \sum_{m=0}^{\infty}
    (-1)^m \ket{\sqrt{2}u e^{i\theta},m}\bra{\sqrt{2}u e^{i\theta},m},
\end{equation}
where the state $\ket{\sqrt{2}u e^{i\theta},m}$ is a particular case of the general state, $\ket{z,m}=\exp(z\hat{a}^\dagger) \ket{m}$.
Then the Wigner phase distribution for the field state $\hat{\rho}_\psi$ is given by,
\begin{equation}\label{Ptheta}
    P_\psi^\text{W} (\theta) = \text{Tr}\left[ \hat{\rho}_\psi \hat{\rho}_\text{W} (\theta) \right].
\end{equation}
Given that the states $\ket{z,m}$ are not orthogonal, but that the Wigner phase operator $\hat{\rho}_\text{W}(\theta)$
obeys the completeness relation,
\begin{equation*}
    \int_0^{2\pi} \hat{\rho}_\text{W}(\theta)\, d\theta = 1,
\end{equation*}
the proper way to understand this formalism is, as an operator valued measure (OVM) on the single-mode field Hilbert space.

OVM based approach to the quantum phase problem has been considered in the past \cite{ShapShep91, Gou04}. The
Shapiro-Shepard \cite{ShapShep91} approach results in the Susskind-Glogower phase operator as the result of a maximum
likelihood estimation on quantum phase measurement, and they
succeed in showing that the so-constructed OVM has similar measurement statistics as that of the Pegg-Barnett hermitian operator.

In this paper, we find the eigenstates of the Wigner phase OVM (WP-OVM), which happen to be rotated quadrature cat states in phase space.
Then we construct the operator satisfying the boson commutation relation with the WP-OVM, which not 
surprisingly, is seen not to be the usual number operator, in view of the fact that the WP-OVM is not
a projective operator. The construction of these two operators, satisfying the boson commutation relation,
in the standard oscillator Hilbert space, completes our program of devising a Wigner distribution based operator formalism for the Wigner Phase Distribution.

The latter section is devoted to addressing the problem of non-positivity of the WP-OVM; the non-positivity being inherited from the potential negativity of the Wigner function itself. We show how a positive OVM can be consistently defined by adopting the view that phase measurements are
 coarse-grained in phase space, i.e., by using a suitable ``filter'' function to
mimic the measurement process. For a suitable ``filter'' function, this leads to a positive OVM based on the familiar Q-distribution. This falls in line with earlier observations regarding the relation between operational phase measurement schemes and the Q-distribution \cite{FreSch93, LeoPau93, LeoPau93A, LeoVacPaul95}.

\section{Wigner phase operator-valued measure formalism}

Before embarking on the important task of constructing the eigenspace of the WP-OVM, we note a very useful property, viz.,
\begin{equation}\label{rhow_theta}
    \hat{\rho}_W(\theta)=e^{i \hat{N} \theta} \hat{\rho}_W(0) e^{-i \hat{N} \theta},
\end{equation}
where $\hat{N}=\hat{a}^\dagger \hat{a}$ is the usual number operator of the field mode. Hence we see the essential connection between
rotations on the phase plane and the phase of a given field state; it can be shown that this property is also shared by the Pegg-Barnett operator, and can be used to establish a weak equivalence between the WP-OVM and 
the Pegg-Barnett formalism \cite{SubViv11}.
This property also holds for other operators as pointed out by Vaccaro \cite{Vacccomarx2012}, and so need not qualify as a sufficient criterion for a phase operator. Nevertheless, it could be a necessary condition.

\subsection{Eigenspace of WP-OVM}

Armed with \eqref{rhow_theta}, we need only consider eigenstates $\ket{\lambda}$ such that,
\begin{equation}\label{rho0_eigeneqn}
    \hat{\rho}_W(0) \ket{\lambda}=\lambda \ket{\lambda},
\end{equation}
so that $e^{i \hat{N} \theta} \ket{\lambda}$ are the eignstates of $\hat{\rho}_W(\theta)$.

Projecting the eigen-equation \eqref{rho0_eigeneqn} onto the position basis $\bra{x}$, and using the $x$-representation
of $\hat{\rho}_W(0)$ viz.,
\begin{equation*}
    \hat{\rho}_W(0) = \frac{1}{2 \pi} \int_0^{\infty} \sqrt{2}x'' ~dx''
        \int_{-\infty}^{\infty}dx'~ \ket{-x'+x''} \bra{x'+x''},
\end{equation*}
we obtain the integral equation,
\begin{equation*}
    \int_{-x'}^{\infty}(x+x')\, f_\lambda(x)\, dx = 4\pi \lambda\, f_\lambda(x')
\end{equation*}
for the $x$-representation, $f_\lambda(x)=\langle x \vert \lambda \rangle$ of the eigenvector $\ket{\lambda}$. Using the technique of
taking derivatives under an integral, the above integral equation can be cast into the differential equation,
\begin{equation*}
    4\pi\lambda~\partial_x^2 f_\lambda(x)=f_\lambda(-x).
\end{equation*}
An elegant way to solve such an equation is to first note that it only connects functions of same parity, which suggests expressing the
equation in the following form,
\begin{equation*}
    4\pi\lambda~ \bra{x} \hat{p}^2 \ket{\lambda} = -\langle{x}\vert \hat{\mathcal{P}} \vert \lambda \rangle,
\end{equation*}
where, $\hat{\mathcal{P}} = \int dx \ket{-x}\bra{x}$ is the parity operator, and $\hat{p} = -i\partial_x$ is the usual momentum operator.
This now implies,
\begin{equation*}
    \hat{p}^2 \ket{\lambda} = \frac{-1}{4\pi\lambda} \hat{\mathcal{P}} \vert \lambda \rangle.
\end{equation*}
Since momentum eigenstates do not have definite parity, this equation can only be satisfied if $\ket{\lambda}$ are definite-parity
superpositions of momentum eigenstates. Careful thought shows that the exact form is,
\begin{equation}
    \ket{\lambda} = \frac{\sqrt{\pi}}{(4 \pi |\lambda|)^{3/4}} \left( \left\vert \frac{1}{\sqrt{4 \pi |\lambda|}} \right\rangle_{\hat{p}}-
        \frac{\lambda}{|\lambda|} \left\vert \dfrac{-1}{\sqrt{4 \pi |\lambda|}} \right\rangle_{\hat{p}} \right),
\end{equation}
where the kets are momentum eigenkets, as indicated by the subscript. This actually represents two disconnected families of
even (for negative $\lambda$) and odd (for positive $\lambda$) eigenstates of $\hat{\rho}_W(0)$ satisfying,
\begin{equation*}
    \hat{\rho}_W(0) \ket{\pm \lambda} = \pm \lambda \ket{\pm \lambda}.
\end{equation*}
It is easy to show that these states furnish an orthonormal basis spanning the Hilbert space, i.e.,
\begin{equation*}
\begin{split}
    & \langle \lambda \vert \lambda' \rangle = \delta(\lambda-\lambda') \\
    & \int_{-\infty}^\infty d\lambda~ \ket{\lambda} \bra{\lambda} = 1.
\end{split}
\end{equation*}

The Canonical conjugate for the Wigner phase operator, as worked out in Appendix(\ref{apencco}),
given by 
\begin{equation*}
    \hat{\eta}_W (0) = -\pi \left(\hat{p}^3 \hat{x} +\hat{x}\hat{p}^3 \right)\hat{\mathcal{P}},
\end{equation*}
is not compatible with the notion of projective phase measurements and therefore suggests that the Wigner
phase operator is not a projector.

\section{Coarse-graining and the positive Q phase OVM}

Given that the Wigner phase operator is indeed not a true projector, and since it exhibits the completeness property of
a set of non-orthogonal measurement operators, one is forced to view them (taken as a continuum of operators for each value of $\theta$)
as an operator valued measure. But the serious trouble with such
an interpretation is the well known non-positivity of the operator \cite{SubViv11,GarKni92}. This section is devoted to offering a particular
solution to this problem, resulting in a positive operator valued measure (POVM).

The non-positivity of the Wigner phase OVM can be traced back to the non-positivity of the postulated definition of the Wigner function
for the phase state which has been noted previously \cite{GarKni92,SubViv11}. It is well known that the degree of negativity of a Wigner
function, the so-called non-classicality depth, is operationally equal to the minimum number of thermal photons to be added to the state
so as to make its Wigner function fully positive \cite{Lee91}.

Hence, if we take the view that experimentally realized phase measurements are in some sense classical, then it could be that one is
able to recover a positive operator valued measure to describe the situation.
So here we suggest that phase measurements are oblivious to the connected patches in phase space of area atmost $\mathcal{O}(\hbar^2)$
where the Wigner phase distribution can be negative; i.e, that classical phase measurements are a coarse-grained
sampling \cite{KofBru07,GelHar07}.

That is, we postulate, instead of the phase probability distribution \eqref{Ptheta} $P_\rho(\theta) = \int W_\rho (\xi)W_\theta(\xi) d^2 \xi$
with the Wigner function $W_\theta$ of the phase operator, a coarse-grained phase probability distribution which corresponds to a
coarse-grained Wigner function for the phase operator,
\begin{equation*}
    W_\theta^\prime(\xi) = \int W_\theta(\xi-\xi') g(\xi') d^2 \xi',
\end{equation*}
where $g$ is the ``filter'' function in phase space that smears out regions of negativity in $W_\theta$.

Now we need to fix the filter function. To do that, we first note that the phase measurement scheme, even though postulated to be classical,
is assumed to be unbiased, i.e., that the measurement process does not introduce any phase-sensitive information in the outcome; this implies
$g(\xi')=g(\vert \xi'\vert)$. Next we demand that the phase measurement be optimal, in the sense that it does not introduce any additional
classical noise. This implies that the optimal $g$ is the one which has a support in phase space comparable to the support of a minimum
uncertainty state, but since it also has to be symmetric, the only possibility is,
$g(\xi') =\exp\left(-\frac{\vert \xi' \vert^2}{2}\right)$. In general, the fact that a particular form of Gaussian smoothing of
the Wigner distribution leads to the Q-distribution is very well known \cite{Lee95,LeoPau93A,Leon97}, and follows rather trivially from the
Cahill-Glauber continuum of $s$-ordered distribution functions \cite{CahGlau69}. As a passing remark, we note that this optimal
coarse-graining corresponds to the addition of a minimum amount of quantum noise as prescribed by the pioneering work of Caves on 
the quantum limits to signal amplification \cite{Cav82}.

By straightforward computation, we find that such a choice in fact defines a new phase probability distribution
based on the Husimi Q-distribution,
\begin{equation*}
    P_\rho^Q (\theta) = \int Q_\rho (\xi) Q_\theta(\xi) d^2 \xi,
\end{equation*}
so that this can be interpreted as the phase probability distribution arising from the radial integration of the $Q_\rho$. Here,
$Q_\theta$ is the angularly peaked Q-distribution satisfying the property,
\begin{equation*}
    \int_0^{2\pi} Q_\rho (r \cos \theta',r \sin \theta')Q_\theta (r \cos \theta-\theta',r \sin \theta-\theta') d\theta'
        = Q_\rho(r \cos \theta, r \sin \theta).
\end{equation*}

\subsection{Properties of the Q phase operator}

Following a procedure similar to the one adopted in the construction of the operator form of the radially integrated Wigner phase
distribution \cite{SubViv11}, we readily find an operator corresponding to the radially integrated Q phase distribution, viz.,
\begin{equation}\label{rhoQ_expression}
    \hat{\rho}_Q (\theta) = \frac{1}{2\pi} \sum_{n,m=0}^\infty \Gamma\left(1+\frac{n+m}{2}\right)
        \frac{e^{i(n-m)\theta}}{\sqrt{n!m!}} \ket{n}\bra{m}.
\end{equation}

From this expression, we calculate the Q phase probability distribution for Fock states,
\begin{equation*}
    P_{\ket{n}\bra{n}}^Q (\theta) = \frac{1}{2\pi},
\end{equation*}
a uniform distribution, as expected of a meaningful phase operator \cite{PegBar89}. Further, for a coherent state we get that,
\begin{equation*}
    P_{\ket{\alpha}\bra{\alpha}}^Q (\theta) = \frac{e^{-\vert\alpha\vert^2}}{2\pi}\left[ 1+\sqrt{\pi}\Re\left(\alpha e^{-i\theta}
        e^{-\frac{i\theta}{2} x(\theta)}\right)+ 2\sqrt{\pi}x(\theta) e^{-\frac{i\theta}{2}x(\theta)}\text{erf}\left(x(\theta)\right)\right],
\end{equation*}
where we have defined the rotated quadrature $x(\theta)=\Re(\alpha e^{-i\theta})$; from the properties of the error function, it
follows that this is a positive phase probability distribution.

\noindent It can also be shown that the Q-phase operator is complete, in the sense that,
\begin{equation}\label{rhoQ_complete}
    \int_0^{2\pi} \hat{\rho}_Q (\theta)\, d\theta = \hat{1}.
\end{equation}
In principle, one could evaluate the eigenstates of the Q-phase operator - this leads to a Fredholm integral equation
with a Gaussian kernel, for the state represented in the coherent state basis. The resulting solution
appeared to be too tedious and not very instructive, therefore we will not present it here.

\begin{figure}[h]
\centering
\includegraphics[scale=0.6]{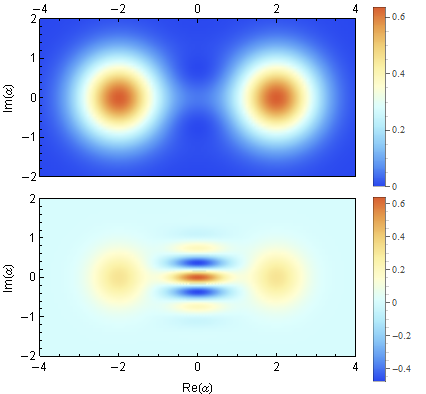}\hfill
\includegraphics[scale=0.5]{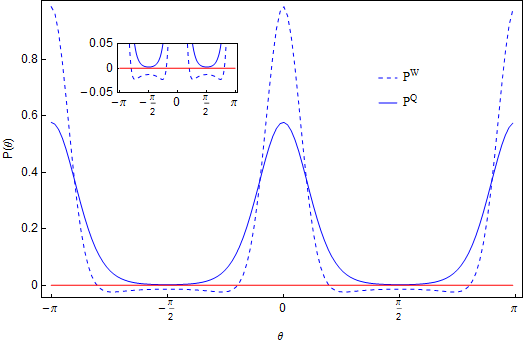}
\caption{\label{fig:cat_state}Figure shows the Q (left panel top) and Wigner (left panel bottom) distributions of the even cat state 
with $\gamma =2$ (see text for details); regions of negative Wigner function is visible, owing to interference in phase space. 
The right panel shows the Wigner and Q phase probability distributions,
clearly depicting the lack of negativity in the Q phase probability, in contrast to the Wigner phase; inset shows a zoom of the 
region of interest.}
\end{figure}

As an illustrative application, we consider the phase probability distribution of an even cat state,
\begin{equation*}
	\ket{\psi} = \frac{\ket{\gamma}+\ket{-\gamma}}{\sqrt{2(1-\exp(-2\vert \gamma \vert^2))}},
\end{equation*}
where $\ket{\gamma}$ is the coherent state with complex amplitude $\gamma$.
It is well known that the Wigner phase distribution of this state is in general negative \cite{GarKni92}, and thus proves to be difficult to
interpret classically. In view of the coarse-graining in phase space that was employed herein to construct the Q phase operator
from its Wigner counterpart, we expect this negativity to disappear. \fref{fig:cat_state} shows the phase space distributions and
the associated phase probability distributions, clearly showing that the Q-phase probability distribution is positive.

\subsection{Q phase operator as a POVM}

Very much like the diagonal resolution \eqref{rhow_OVM}, one can arrive at a diagonal form for $\hat{\rho}_Q (\theta)$. 
For this case of the Q-phase distribution, this can be most easily deduced by expressing the phase probability distribution
as a radial integral over a coherent state expectation of the relevant state, i.e.,
\begin{equation*}
    P_{\psi}^Q (\theta) = \frac{1}{\pi}\int_0^\infty dr r\; \bra{r e^{i\theta}} \rho_\psi \ket{r e^{i\theta}},
\end{equation*}
where $\ket{r e^{i\theta}}$ is the phase rotated coherent state $\exp(i\hat{N}\theta) \ket{r}$. From this we readily identify,
\begin{equation}\label{rhoQ_diagonal}
    \hat{\rho}_Q (\theta) = \frac{1}{\pi}\int_0^\infty dr r\; \ket{r e^{i\theta}} \bra{r e^{i\theta}},
\end{equation}
which is the required diagonal representation of the operator.

Together with the completeness result \eqref{rhoQ_complete}, and the fact that the Q-phase distribution is
positive by construction (either as the operator corresponding to the radially integrated Q-distribution, or equivalently
as a coarse-graining of the Wigner phase operator adopted above), this shows that the Q-phase operator \eqref{rhoQ_expression}
is a proper POVM. 

This construction can be directly compared to a previous investigation on the same theme \cite{LeoVacPaul95}, where the authors
succeed in starting from the radially integrated Q-distribution, and then use it to arrive at the Wigner phase distribution and
a more general $s$-parametrized phase distribution. In fact, there, and previously \cite{LaiHaus89,LeoPau93}, it is shown that
experimentally relevant phase measurement schemes with a strong local oscillator are in fact equivalent to the radially integrated
Q-distribution, for which we have constructed here, an operator representation, as well as a POVM resolution. On the other hand, 
our construction can be compared with a purely operational result obtained in \cite{BusLah95}, obtained as the result of a
necessary constraint required of any phase measurement scheme - invariant to phase shifts by integer multiples of $2\pi$.

\subsection{Dilation of the Q phase POVM}

It is well known, owing to the Gelfand-Naimark theorem, that POVM's on a subsystem codify the action of a von-Neumann projector
on a bigger system, of which the subsystem is a part. This means that a POVM can be understood as the ``dilation'' of some projector
into a larger Hilbert space, followed by the action of a unitary operation, succeeded by
the reduction of the resulting state back into the original Hilbert space. Quite obviously, this dilation is ambiguous. But through
this ambiguity enters the freedom, and hence the generalization, required to specify operational measurement schemes within quantum theory.

In the case of phase measurements, one plausible operational definition of the measurement process is the interference of the given state under consideration with a known
coherent state at a beam-splitter. In this last part of the paper, we would like to show how the Q phase POVM can be understood as the
result of a dilation which captures this idea.

Consider the POVM generated by,
\begin{equation}\label{Pi_definition}
 \hat{\Pi}_\tau (\beta) \equiv \text{Tr}_b \left[ \hat{U}_\tau \left( \ket{\psi}\otimes \ket{\beta} \right)
 \left( \bra{\beta}\otimes \bra{\psi} \right) \hat{U}_{\tau}^{\dagger}
 \right],
\end{equation}
where $\ket{\beta}$ is a coherent state in Hilbert space $\mathcal{H}_b$, $\ket{\psi} \in \mathcal{H}_a$ is associated to the projector
$\ket{\psi}\bra{\psi}$, and $\hat{U}_\tau = \exp \left[i\tau \left(\hat{a}^\dagger \hat{b} + \hat{a} \hat{b}^\dagger \right)\right]$ is the
propagator for a beam-splitter with transmissivity $\cos^2 2\tau$.

\eqref{Pi_definition} can be explicitly evaluated, assuming a P-representation $P(\alpha)$ for the projector $\ket{\psi}\bra{\psi}$, giving,
\begin{equation*}
 \hat{\Pi}_\tau (\beta) = \int_0^\infty \int_0^{2\pi} \frac{dr\, d\phi}{\pi}\, r\; P(r,\phi) 
			  \ket{re^{i\phi}\cos \tau +i\beta \sin \tau}\bra{re^{i\phi}\cos \tau +i\beta \sin \tau},
\end{equation*}
where the state within the integral is a coherent state of the argument. If this operational procedure is to be represented by the Q phase POVM
\eqref{rhoQ_diagonal}, then we must have that: (a) $P(r,\phi) = \delta(\phi -\theta)$, an angularly peaked delta function,
and (b) $\tau \rightarrow 0$, corresponding to a transmissivity of near unity for the state $\ket{\psi}$, implying
that the ancilla coherent state must have $\beta \rightarrow \infty$ (for the practical reason of obtaining any signal at all).

Thus, for an operational setting defined by interference at a beam-splitter with a coherent state, the Q phase POVM has the dilation,
\begin{equation*}
 \hat{\rho}_Q (\theta) = \lim_{\tau\rightarrow 0, \beta \rightarrow \infty} 
	\text{Tr}_b \left[ \hat{U}_\tau \left(\int \frac{dr\, d\phi}{\pi}r \delta(\phi-\theta) \ket{re^{i\phi},\beta}\bra{re^{i\phi},\beta}\right) \hat{U}_\tau^\dagger \right].
\end{equation*}

\section{Conclusion}

We have brought to a conclusion, the program of devising an operator formalism for the radially integrated 
phase space distributions. 
In the process, we discover that the natural setting of such an operator formalism is that of an operator valued measure on the single
mode Hilbert space.

Although the radially integrated Wigner phase operator is not positive, we show that
by assuming phase measurements as a phase-unbiased, but quantum-limited, smoothing of the Wigner phase operator, one can indeed construct a
positive operator valued measure. This object is the operator corresponding to the radially integrated Q-distribution,
which has been previously shown in several contexts to be related to an operational description of quantum optical phase.

Taking the operational approach further, we are able to show that the so-obtained Q phase POVM can be dilated into a von Neumann projector
in a two-mode Hilbert space. By demanding that the operation involved is interference at a beam-splitter with a coherent state, we are naturally
led to a projector which has a well-defined phase. But quite obviously, this is one among an infinite possible dilations of the POVM.

\section*{Acknowledgments}

One of the authors (VS) would like to thank Prof. Myungshik Kim at the Dept. of Physics, Imperial College London for his constructive comments
and perusal of the manuscript.

\bibliography{ref_wigner2}

\begin{thebibliography}{33}
\expandafter\ifx\csname natexlab\endcsname\relax\def\natexlab#1{#1}\fi
\expandafter\ifx\csname bibnamefont\endcsname\relax
  \def\bibnamefont#1{#1}\fi
\expandafter\ifx\csname bibfnamefont\endcsname\relax
  \def\bibfnamefont#1{#1}\fi
\expandafter\ifx\csname citenamefont\endcsname\relax
  \def\citenamefont#1{#1}\fi
\expandafter\ifx\csname url\endcsname\relax
  \def\url#1{\texttt{#1}}\fi
\expandafter\ifx\csname urlprefix\endcsname\relax\def\urlprefix{URL }\fi
\providecommand{\bibinfo}[2]{#2}
\providecommand{\eprint}[2][]{\url{#2}}

\bibitem[{\citenamefont{Barnett and Vaccaro}(2007)}]{BarVac07}
\bibinfo{author}{\bibfnamefont{S.~M.} \bibnamefont{Barnett}} \bibnamefont{and}
  \bibinfo{author}{\bibfnamefont{J.~A.} \bibnamefont{Vaccaro}},
  \emph{\bibinfo{title}{The Quantum Phase Operator}}
  (\bibinfo{publisher}{Taylor $\&$ Francis, New York}, \bibinfo{year}{2007}).

\bibitem[{\citenamefont{Nieto}(1993)}]{Nieto93}
\bibinfo{author}{\bibfnamefont{M.~M.} \bibnamefont{Nieto}},
  \bibinfo{journal}{Phys. Scr.} \textbf{\bibinfo{volume}{T48}},
  \bibinfo{pages}{5} (\bibinfo{year}{1993}).

\bibitem[{\citenamefont{Dirac}(1927)}]{Dirac27}
\bibinfo{author}{\bibfnamefont{P.~A.~M.} \bibnamefont{Dirac}},
  \bibinfo{journal}{Proc. R. Soc. London} \textbf{\bibinfo{volume}{A114}},
  \bibinfo{pages}{243} (\bibinfo{year}{1927}).

\bibitem[{\citenamefont{Stenholm}(1993)}]{Sten93}
\bibinfo{author}{\bibfnamefont{S.}~\bibnamefont{Stenholm}},
  \bibinfo{journal}{Phys. Scr.} \textbf{\bibinfo{volume}{T48}},
  \bibinfo{pages}{77} (\bibinfo{year}{1993}).

\bibitem[{\citenamefont{Louisell}(1963)}]{Loui63}
\bibinfo{author}{\bibfnamefont{W.~H.} \bibnamefont{Louisell}},
  \bibinfo{journal}{Phys. Lett.} \textbf{\bibinfo{volume}{7}},
  \bibinfo{pages}{60} (\bibinfo{year}{1963}).

\bibitem[{\citenamefont{Susskind and Glogower}(1964)}]{SusGlo64}
\bibinfo{author}{\bibfnamefont{L.}~\bibnamefont{Susskind}} \bibnamefont{and}
  \bibinfo{author}{\bibfnamefont{J.}~\bibnamefont{Glogower}},
  \bibinfo{journal}{Physics} \textbf{\bibinfo{volume}{1}}, \bibinfo{pages}{49}
  (\bibinfo{year}{1964}).

\bibitem[{\citenamefont{Carruthers and Nieto}(1965)}]{CarNie65}
\bibinfo{author}{\bibfnamefont{P.}~\bibnamefont{Carruthers}} \bibnamefont{and}
  \bibinfo{author}{\bibfnamefont{M.~M.} \bibnamefont{Nieto}},
  \bibinfo{journal}{Phys. Rev. Lett.} \textbf{\bibinfo{volume}{14}},
  \bibinfo{pages}{387} (\bibinfo{year}{1965}).

\bibitem[{\citenamefont{Carruthers and Nieto}(1968)}]{CarNie68}
\bibinfo{author}{\bibfnamefont{P.}~\bibnamefont{Carruthers}} \bibnamefont{and}
  \bibinfo{author}{\bibfnamefont{M.~M.} \bibnamefont{Nieto}},
  \bibinfo{journal}{Rev. Mod. Phys.} \textbf{\bibinfo{volume}{40}},
  \bibinfo{pages}{411} (\bibinfo{year}{1968}).

\bibitem[{\citenamefont{Noh et~al.}(1992)\citenamefont{Noh, Fougeres, and
  Mandel}}]{NFM92}
\bibinfo{author}{\bibfnamefont{J.~W.} \bibnamefont{Noh}},
  \bibinfo{author}{\bibfnamefont{A.}~\bibnamefont{Fougeres}}, \bibnamefont{and}
  \bibinfo{author}{\bibfnamefont{L.}~\bibnamefont{Mandel}},
  \bibinfo{journal}{Phys. Rev. A} \textbf{\bibinfo{volume}{45}},
  \bibinfo{pages}{424} (\bibinfo{year}{1992}).

\bibitem[{\citenamefont{Barnett and Pegg}(1986)}]{PegBar86}
\bibinfo{author}{\bibfnamefont{S.~M.} \bibnamefont{Barnett}} \bibnamefont{and}
  \bibinfo{author}{\bibfnamefont{D.~T.} \bibnamefont{Pegg}},
  \bibinfo{journal}{J. Phys. A: Math. Gen.} \textbf{\bibinfo{volume}{19}},
  \bibinfo{pages}{3849} (\bibinfo{year}{1986}).

\bibitem[{\citenamefont{Pegg and Barnett}(1989)}]{PegBar89}
\bibinfo{author}{\bibfnamefont{D.~T.} \bibnamefont{Pegg}} \bibnamefont{and}
  \bibinfo{author}{\bibfnamefont{S.~M.} \bibnamefont{Barnett}},
  \bibinfo{journal}{Phys. Rev. A} \textbf{\bibinfo{volume}{39}},
  \bibinfo{pages}{1665} (\bibinfo{year}{1989}).

\bibitem[{\citenamefont{Barnett and Dalton}(1993)}]{BarnDalt93}
\bibinfo{author}{\bibfnamefont{S.~M.} \bibnamefont{Barnett}} \bibnamefont{and}
  \bibinfo{author}{\bibfnamefont{B.~J.} \bibnamefont{Dalton}},
  \bibinfo{journal}{Phys. Scr.} \textbf{\bibinfo{volume}{T48}},
  \bibinfo{pages}{13} (\bibinfo{year}{1993}).

\bibitem[{\citenamefont{Schleich et~al.}(1992)\citenamefont{Schleich, Bandilla,
  and Paul}}]{SchBan92}
\bibinfo{author}{\bibfnamefont{W.}~\bibnamefont{Schleich}},
  \bibinfo{author}{\bibfnamefont{A.}~\bibnamefont{Bandilla}}, \bibnamefont{and}
  \bibinfo{author}{\bibfnamefont{H.}~\bibnamefont{Paul}},
  \bibinfo{journal}{Phys. Rev. A} \textbf{\bibinfo{volume}{45}},
  \bibinfo{pages}{6652} (\bibinfo{year}{1992}).

\bibitem[{\citenamefont{Garraway and Knight}(1992)}]{GarKni92}
\bibinfo{author}{\bibfnamefont{B.~M.} \bibnamefont{Garraway}} \bibnamefont{and}
  \bibinfo{author}{\bibfnamefont{P.~L.} \bibnamefont{Knight}},
  \bibinfo{journal}{Phys. Rev. A} \textbf{\bibinfo{volume}{46}},
  \bibinfo{pages}{5346} (\bibinfo{year}{1992}).

\bibitem[{\citenamefont{Tanas et~al.}(1993)\citenamefont{Tanas, Miranowicz, and
  Gantsog}}]{TanGan93}
\bibinfo{author}{\bibfnamefont{R.}~\bibnamefont{Tanas}},
  \bibinfo{author}{\bibfnamefont{A.}~\bibnamefont{Miranowicz}},
  \bibnamefont{and} \bibinfo{author}{\bibfnamefont{T.}~\bibnamefont{Gantsog}},
  \bibinfo{journal}{Phys. Scr.} \textbf{\bibinfo{volume}{T48}},
  \bibinfo{pages}{53} (\bibinfo{year}{1993}).

\bibitem[{\citenamefont{Subeesh and Sudhir}(2011)}]{SubViv11}
\bibinfo{author}{\bibfnamefont{T.}~\bibnamefont{Subeesh}} \bibnamefont{and}
  \bibinfo{author}{\bibfnamefont{V.}~\bibnamefont{Sudhir}},
  \bibinfo{journal}{J. Mod. Opt.} \textbf{\bibinfo{volume}{58}},
  \bibinfo{pages}{761} (\bibinfo{year}{2011}).

\bibitem[{\citenamefont{Shapiro and Shepard}(1991)}]{ShapShep91}
\bibinfo{author}{\bibfnamefont{J.~H.} \bibnamefont{Shapiro}} \bibnamefont{and}
  \bibinfo{author}{\bibfnamefont{S.~R.} \bibnamefont{Shepard}},
  \bibinfo{journal}{Phys. Rev. A} \textbf{\bibinfo{volume}{43}},
  \bibinfo{pages}{3795} (\bibinfo{year}{1991}).

\bibitem[{\citenamefont{Gour et~al.}(2004)\citenamefont{Gour, Khanna, and
  Revzen}}]{Gou04}
\bibinfo{author}{\bibfnamefont{G.}~\bibnamefont{Gour}},
  \bibinfo{author}{\bibfnamefont{F.~C.} \bibnamefont{Khanna}},
  \bibnamefont{and} \bibinfo{author}{\bibfnamefont{M.}~\bibnamefont{Revzen}},
  \bibinfo{journal}{Phys. Rev. A} \textbf{\bibinfo{volume}{69}},
  \bibinfo{pages}{014101} (\bibinfo{year}{2004}).

\bibitem[{\citenamefont{Freyberger et~al.}(1993)\citenamefont{Freyberger,
  Vogel, and Schleich}}]{FreSch93}
\bibinfo{author}{\bibfnamefont{M.}~\bibnamefont{Freyberger}},
  \bibinfo{author}{\bibfnamefont{K.}~\bibnamefont{Vogel}}, \bibnamefont{and}
  \bibinfo{author}{\bibfnamefont{W.~P.} \bibnamefont{Schleich}},
  \bibinfo{journal}{Phys. Lett. A} \textbf{\bibinfo{volume}{176}},
  \bibinfo{pages}{41} (\bibinfo{year}{1993}).

\bibitem[{\citenamefont{Leonhardt and Paul}(1993{\natexlab{a}})}]{LeoPau93}
\bibinfo{author}{\bibfnamefont{U.}~\bibnamefont{Leonhardt}} \bibnamefont{and}
  \bibinfo{author}{\bibfnamefont{H.}~\bibnamefont{Paul}},
  \bibinfo{journal}{Phys. Rev. A} \textbf{\bibinfo{volume}{47}},
  \bibinfo{pages}{R2460} (\bibinfo{year}{1993}{\natexlab{a}}).

\bibitem[{\citenamefont{Leonhardt and Paul}(1993{\natexlab{b}})}]{LeoPau93A}
\bibinfo{author}{\bibfnamefont{U.}~\bibnamefont{Leonhardt}} \bibnamefont{and}
  \bibinfo{author}{\bibfnamefont{H.}~\bibnamefont{Paul}},
  \bibinfo{journal}{Phys. Rev. A} \textbf{\bibinfo{volume}{48}},
  \bibinfo{pages}{4598} (\bibinfo{year}{1993}{\natexlab{b}}).

\bibitem[{\citenamefont{Leonhardt et~al.}(1995)\citenamefont{Leonhardt,
  Vaccaro, Boehmer, and Paul}}]{LeoVacPaul95}
\bibinfo{author}{\bibfnamefont{U.}~\bibnamefont{Leonhardt}},
  \bibinfo{author}{\bibfnamefont{J.~A.} \bibnamefont{Vaccaro}},
  \bibinfo{author}{\bibfnamefont{B.}~\bibnamefont{Boehmer}}, \bibnamefont{and}
  \bibinfo{author}{\bibfnamefont{H.}~\bibnamefont{Paul}},
  \bibinfo{journal}{Phys. Rev. A} \textbf{\bibinfo{volume}{51}},
  \bibinfo{pages}{84} (\bibinfo{year}{1995}).

\bibitem[{\citenamefont{Vaccaro}(2012)}]{Vacccomarx2012}
\bibinfo{author}{\bibfnamefont{J.~A.} \bibnamefont{Vaccaro}},
  \bibinfo{journal}{arXiv:1209.0870 [quant-ph]}  (\bibinfo{year}{2012}).

\bibitem[{\citenamefont{Lee}(1991)}]{Lee91}
\bibinfo{author}{\bibfnamefont{C.~T.} \bibnamefont{Lee}},
  \bibinfo{journal}{Phys. Rev. A} \textbf{\bibinfo{volume}{44}},
  \bibinfo{pages}{R2775} (\bibinfo{year}{1991}).

\bibitem[{\citenamefont{Kofler and Brukner}(2007)}]{KofBru07}
\bibinfo{author}{\bibfnamefont{J.}~\bibnamefont{Kofler}} \bibnamefont{and}
  \bibinfo{author}{\bibfnamefont{C.}~\bibnamefont{Brukner}},
  \bibinfo{journal}{Phys. Rev. Lett.} \textbf{\bibinfo{volume}{99}},
  \bibinfo{pages}{180403} (\bibinfo{year}{2007}).

\bibitem[{\citenamefont{Gell-Mann and Hartle}(2007)}]{GelHar07}
\bibinfo{author}{\bibfnamefont{M.}~\bibnamefont{Gell-Mann}} \bibnamefont{and}
  \bibinfo{author}{\bibfnamefont{J.~B.} \bibnamefont{Hartle}},
  \bibinfo{journal}{Phys. Rev. A} \textbf{\bibinfo{volume}{76}},
  \bibinfo{pages}{022104} (\bibinfo{year}{2007}).

\bibitem[{\citenamefont{Lee}(1995)}]{Lee95}
\bibinfo{author}{\bibfnamefont{H.-W.} \bibnamefont{Lee}},
  \bibinfo{journal}{Phys. Rep.} \textbf{\bibinfo{volume}{259}},
  \bibinfo{pages}{147} (\bibinfo{year}{1995}).

\bibitem[{\citenamefont{Leonhardt}(1997)}]{Leon97}
\bibinfo{author}{\bibfnamefont{U.}~\bibnamefont{Leonhardt}},
  \emph{\bibinfo{title}{Measuring the Quantum State of Light}}
  (\bibinfo{publisher}{Cambridge University Press}, \bibinfo{year}{1997}).

\bibitem[{\citenamefont{Cahill and Glauber}(1969)}]{CahGlau69}
\bibinfo{author}{\bibfnamefont{K.~E.} \bibnamefont{Cahill}} \bibnamefont{and}
  \bibinfo{author}{\bibfnamefont{R.~J.} \bibnamefont{Glauber}},
  \bibinfo{journal}{Phys. Rev.} \textbf{\bibinfo{volume}{177}},
  \bibinfo{pages}{5} (\bibinfo{year}{1969}).

\bibitem[{\citenamefont{Caves}(1982)}]{Cav82}
\bibinfo{author}{\bibfnamefont{C.~M.} \bibnamefont{Caves}},
  \bibinfo{journal}{Phys. Rev. D} \textbf{\bibinfo{volume}{26}},
  \bibinfo{pages}{1817} (\bibinfo{year}{1982}).

\bibitem[{\citenamefont{Lai and Haus}(1989)}]{LaiHaus89}
\bibinfo{author}{\bibfnamefont{Y.}~\bibnamefont{Lai}} \bibnamefont{and}
  \bibinfo{author}{\bibfnamefont{H.~A.} \bibnamefont{Haus}},
  \bibinfo{journal}{Quantum Opt.} \textbf{\bibinfo{volume}{1}},
  \bibinfo{pages}{99} (\bibinfo{year}{1989}).

\bibitem[{\citenamefont{Busch et~al.}(1995)\citenamefont{Busch, Grabowski, and
  Lahti}}]{BusLah95}
\bibinfo{author}{\bibfnamefont{P.}~\bibnamefont{Busch}},
  \bibinfo{author}{\bibfnamefont{M.}~\bibnamefont{Grabowski}},
  \bibnamefont{and} \bibinfo{author}{\bibfnamefont{P.~J.} \bibnamefont{Lahti}},
  \bibinfo{journal}{Ann. Phys.} \textbf{\bibinfo{volume}{237}},
  \bibinfo{pages}{1} (\bibinfo{year}{1995}).

\bibitem[{\citenamefont{D.T.Pegg et~al.}(1990)\citenamefont{D.T.Pegg, Vaccaro,
  and Barnett}}]{PegVacBar90}
\bibinfo{author}{\bibnamefont{D.T.Pegg}},
  \bibinfo{author}{\bibfnamefont{J.}~\bibnamefont{Vaccaro}}, \bibnamefont{and}
  \bibinfo{author}{\bibfnamefont{S.}~\bibnamefont{Barnett}},
  \bibinfo{journal}{J. Mod. Opt.} \textbf{\bibinfo{volume}{37}},
  \bibinfo{pages}{1703} (\bibinfo{year}{1990}).

\end{thebibliography}

\appendix
\section{Canonical conjugate operator}\label{apencco}

Although we have calculated the orthonormal eigenstates of the Wigner phase operator, it does not assert that the operator
implements projective
phase measurements. As has been emphasized above, the proper way is to view our operator as an operator valued measure. To clearly show the
departure of our operator 
from the phase operators that come under the standard von Neumann prescription,
 we now deduce its hypothetical conjugate operator $\hat{\eta}_W$, defined
by the canonical commutation,
\begin{equation*}
    \left[ \hat{\rho}_W,\hat{\eta}_W \right] = i.
\end{equation*}
If such a conjugate operator existed and were to be compatible with the notion of projective phase measurements via $\hat{\rho}_W$, then
the only possibility, as has been amply pursued in the past in the form of the number-phase uncertainty relation \cite{Dirac27,CarNie68},
is for the conjugate operator to be the standard number operator \cite{PegVacBar90}.

One can explicitly determine $\hat{\eta}_W$; to do so, it is sufficient to find the operator canonically conjugate to $\hat{\rho}_W(0)$. The
latter statement follows by taking into account the manifest ambiguity in assigning the orientation of the axes in phase space, and the fact
that $\hat{\rho}_W (\theta)$ is generated from $\hat{\rho}_W (0)$ by rotations in phase space.

If $\ket{\tilde{\lambda}}$ are eigenstates of $\hat{\eta}_W (0)$ corresponding to eigenvalue $\tilde{\lambda}$, then it follows that they
are given by the delta-normalized wavefunction,
\begin{equation*}
    \ket{\tilde{\lambda}} = \frac{1}{\sqrt{2\pi}} \exp\left(i\tilde{\lambda}\hat{\rho}_W (0) \right)
                            \int_{-\infty}^{\infty} \ket{\lambda}\, d\lambda,
\end{equation*}
giving the wavefunction in the $\lambda$-representation, $\bra{\lambda}\tilde{\lambda}\rangle = \frac{1}{\sqrt{2\pi}}
\exp(i\lambda \tilde{\lambda})$. By using the above expression for the eigenstate, one can work out the expression for the conjugate operator,
which comes out to be,
\begin{equation*}
    \hat{\eta}_W (0) = -\pi \left(\hat{p}^3 \hat{x} +\hat{x}\hat{p}^3 \right)\hat{\mathcal{P}}.
\end{equation*}
Since this is not the number operator, it follows that its conjugate $\hat{\rho}_W (0)$ does not implement projective phase measurements.

\end{document}